\documentclass[nofootinbib,letterpaper,aps,amsmath,amsfonts,twocolumn]{revtex4}
\usepackage[mathscr]{eucal}
\newcommand{\td}{\textup{d}}  
\def\Pcm#1{{\mathcal{#1}}}
\newcommand{\del}{\partial}
\def\eqref#1{(\ref{#1})}
\def\er#1{eqn.\eqref{#1}}
\def\nn{\nonumber}
\begin{document}
\title{Universality of correction to L\"uscher term in Polchinski-Strominger effective string theories}
\author{N.~D.~Hari Dass} 
\email{dass@imsc.res.in}
\affiliation{The Institute of Mathematical Sciences, Chennai 600-113, INDIA}
\author{Peter Matlock} 
\email{pwm@imsc.res.in}
\affiliation{The Institute of Mathematical Sciences, Chennai 600-113, INDIA}
\begin{abstract}
  We show, by explicit calculation, that the next correction to the
  universal L\"uscher term in the effective string theories of
  Polchinski and Strominger is also universal.  We find that to this
  order in inverse string-length, the ground-state energy as well as
  the excited-state energies are the same as those given by the
  Nambu-Goto string theory, the difference being that while the
  Nambu-Goto theory is inconsistent outside the critical dimension,
  the Polchinski-Strominger theory is by construction consistent for
  any space-time dimension. Our calculation explicitly avoids the use
  of any field redefinitions as they bring in many other issues that
  are likely to obscure the main points.
\end{abstract}

\maketitle

\section{Introduction}
{\label{intro}}

Fundamental string theories can only be consistently quantised in the
so-called critical dimension which is $D=26$ for bosonic and $D=10$
for supersymmetric theories.  On the other hand string-like defects or
solitons occur in a wide variety of physical circumstances, the most
well-known being vortices in superfluids, the Nielsen-Olesen vortices
of quantum field theories, vortices in Bose-Einstein condensates and
QCD strings.  These objects do clearly exist in dimensions other than
the previously mentioned critical dimensions.  The challenge then is
to find means of consistently quantising such effective string
theories without restriction on the dimension.

Polchinski and Strominger (PS) \cite{PS} indeed showed how to do this.
Their proposal is in spirit very close to that of chiral perturbation
theory \cite{anant}, which is an effective description of QCD at low
energies. While requiring the symmetries of QCD to be maintained, it
is otherwise unconstrained by requirements like polynomial lagrangians
and renormalisability.  Likewise PS advocated including all possible
terms in the action that preserve the constraint and symmetry
structure of string theories.  The action terms they propose are not
polynomial and in fact can become singular for certain string
configurations. However, understood as terms in an effective action,
they are to be used in a \emph{long-string} vacuum, for which the
dominant term in the action is the usual quadratic action. This allows
perturbation in the small parameter $R^{-1}$ where $2\pi R$ is the
length of the (closed) string. In addition, PS dropped terms in the
action which are proportional to equations of motion and constraints,
to appropriate orders in $1/R$.

The plan of the paper is as follows. In the next section we briefly
review the PS scheme to order $R^{-2}$. We then prove, in very general
terms, the absence of additional terms in the action which are of
order $R^{-3}$. This is crucial in establishing the results of this
paper. Using this we carry out the analysis of the spectrum to higher
orders, where we show the absence of order-$R^{-2}$ terms as well
absence of corrections to order-$R^{-3}$ terms.

In our analysis, we have carefully avoided the use of any additional
ingredients such as field redefinitions. Field redefinitions bring
with them a number of new issues like associated changes in measures
and intrinsic arbitrariness.  While we have nothing against these
\emph{per se}, we wish to present an analysis that is not obscured by
them.

\section{Leading-order analysis}
Here, we review the analysis given by Polchinski and Strominger \cite{PS}.
They begin with the action
\begin{eqnarray}
\label{action2}
S &=& \frac{1}{4\pi} \int \td\tau^+ \td\tau^- \bigg\{
 \frac{1}{a^2} \del_+ X^\mu \del_- X_\mu \nn\\
&&\mbox{}+\beta \frac{\del_+^2 X\cdot\del_- X \del_+ X\cdot\del_-^2 X}{(\del_+X\cdot\del_-X)^2} 
+\Pcm{O}(R^{-3})
\bigg\}
.\end{eqnarray}
This action is invariant under the modified conformal transformations
\begin{equation}
\label{modtrans}
\delta_- X^\mu = \epsilon^-(\tau^-)\del_- X^\mu - \frac{\beta a^2}{2}\del_-^2 \epsilon^-(\tau^-)
\frac{\del_+ X^\mu}{\del_+X\cdot\del_-X}
,\end{equation}
(and another; $\delta_+X$ with $+$ and $-$ interchanged)
leading to the energy momentum tensor (which agrees with eqn(11) of \cite{PS}
to the relevant order)
\begin{eqnarray}
\label{fullT__}
T_{--}^{\textup{PS}} &=& -\frac{1}{2a^2}\del_-X\cdot\del_-X
+\frac{\beta}{2L^2}\big(
     L\del_-^2 L-(\del_- L)^2 \nn\\
&+&\del_-X\cdot\del_-X \del_+^2 X\cdot\del_-^2 X -\del_+ L \del_-X\cdot\del_-^2 X 
     \big)
\end{eqnarray}
where we have omitted terms proportional to the leading-order equation 
of motion,
$ \del_+\del_- X^\mu = 0 $ which has the solution
$X^\mu_{\textup{cl}} = e^\mu_+R\tau^+ + e^\mu_- R \tau^-$;
here $e_-^2=e_+^2=0$ and $e_+\cdot e_- = -1/2$.
Fluctuations around the classical solution are denoted by $Y^\mu$, so that
$
X^\mu = X^\mu_{\textup{cl}}+Y^\mu
$.
The energy-momentum tensor in terms of the fluctuation field is then
\begin{equation}
\label{T_2}
T_{--} = -\frac{R}{a^2}e_\cdot\del_-Y -\frac{1}{2a^2}\del_-Y\cdot\del_-Y
-\frac{\beta}{R}e_+\cdot\del_-^3 Y+\ldots
\end{equation}
with the OPE of
$T_{--}(\tau^-)T_{--}(0)$ given by 
\begin{equation}
\frac{D+12\beta}{2(\tau^-)^4}+\frac{2}{(\tau^-)^2}T_{--}
+\frac{1}{\tau^-}\del_-T_{--} 
+\Pcm{O}(R^{-1})
.\end{equation}

It should be noted that due to the 
$-\frac{R}{a^2}e_-\cdot\del_-Y$ term in $T_{--}$, in principle the order-$R^{-2}$
term in the $Y$-$Y$ propagator could contribute. 
It turns out that for the PS field definition it does not. 
The full equation of motion is $E^\mu=0$;
\begin{eqnarray}
\label{modEOM}
E^\mu &=& -\frac{1}{2\pi a^2}\del_{+-}X^\mu
+\frac{\beta}{4\pi}\bigg[
       \del_+^2\big\{
          \frac{\del_-X^\mu(\del_-^2X\cdot\del_+X)}{L^2}\big\} \nn\\
&+&2\del_+\big\{\frac{\del_-X^\mu(\del_+^2X\cdot\del_-X)(\del_-^2X\cdot\del_+X)}{L^3}\big\} 
\nn\\
       &-&\del_-\big\{\frac{\del_+^2X^\mu(\del_-^2X\cdot\del_+X)}{L^2}\big\} 
+\{+\leftrightarrow -\}\bigg]
,\end{eqnarray}
where we have used the notation $L=\del_+X\cdot\del_-X$.
When this equation is restricted to terms linear in $Y^\mu$ we 
get an equation from which the two-point function can be computed;
\begin{equation}
\langle Y^\mu Y^\nu \rangle = -a^2\log(\tau^+\tau^-)\eta^{\mu\nu}
+2\frac{\beta a^4}{R^2}
e_-^{(\mu} e_+^{\nu)}\delta^2(\tau)
\end{equation}

Consequently the potential contribution to the central charge
$\frac{R^2}{a^4} e_-^\mu e_-^\nu \langle Y^\mu Y^\nu \rangle$
vanishes, as $e_-\cdot e_- =0$. This is not always true as can be
checked by redefining the $X^\mu$ field. Of course the total central
charge does not change. One must add the contribution $-26$ from the
ghosts, leading to the total central charge $D+12\beta-26$. Vanishing
of the conformal anomaly thus requires
\begin{equation}
\beta_c = -\frac{D-26}{12}
,\end{equation}
valid for any dimension $D$.

Using standard techniques the spectrum of this effective theory can be
worked out. PS have shown how to do this at the leading order.  We
briefly reproduce their results here in order to set the stage for the
rest of the paper. The Virasoro generators operate on the Fock space
basis provided by 
$\del_-Y^\mu = a\sum_{m=-\infty}^{\infty}\alpha_m^\mu e^{-im\tau^-}$ 
and are given by
\begin{eqnarray}
\label{virasoro}
L_n &=& \frac{R}{a}e_-\cdot\alpha_n
+\frac12\sum_{m=-\infty}^\infty :\alpha_{n-m}\cdot\alpha_m: \nn\\
&+&\frac{\beta_c}{2}\delta_{n}
-\frac{a\beta_c n^2}{R}e_+\cdot\alpha_n + \Pcm{O}(R^{-2})
.\end{eqnarray}

The quantum ground state is $|k,k;0\rangle$ which is also an
eigenstate of $\alpha_0^\mu$ and ${\tilde\alpha}_0^\mu$ with common
eigenvalue $ak^\mu$.  This state is annihilated by all $\alpha_n^\mu$
for positive-definite $n$.  The ground state momentum is
$p^\mu_{\textup{gnd}} = \frac{R}{2a^2}(e_+^\mu+e_-^\mu) + k^\mu$ while
the total rest energy is
\begin{equation}
\label{restenergy}
(-p^2)^{1/2} = \sqrt{
\left(\frac{R}{2a^2}\right)^2
-k^2-\frac{R}{a^2}(e_++e_-)\cdot k } 
.\end{equation}
The physical state conditions $L_0 = {\tilde{L}}_0 = 1$ fix $k$, so that
\begin{equation}
\label{leadingvir}
k^1 = 0,\qquad
k^2+\frac{R}{a^2}(e_++e_-)\cdot k = \frac{(2-\beta_c)}{a^2}
.\end{equation}
The first follows from the periodic boundary condition for the closed string
which gives $e_+^\mu-e_-^\mu = \delta^\mu_1$. Substituting the critical value
$\beta_c = (26-D)/12$ one arrives at
\begin{equation}
(-p^2)^{1/2} = \frac{R}{2a^2}\sqrt{1-\frac{D-2}{12}\left(\frac{2a}{R}\right)^2}
,\end{equation}
which is the precise analog of the result obtained by Arvis for open strings
\cite{Arv}. Expanding this and keeping
only the first correction, one obtains for the static potential
\begin{equation}
V(r) = \frac{R}{2a^2} -\frac{D-2}{12}\frac{1}{R}+\cdots
.\end{equation}
\section{Absence of additional terms at order $R^{-3}$}
\label{sec_absence}
It is of crucial importance for the arguments of this paper that the
next possible candidate term in the action is not $R^{-3}$ order.  PS
have stated without proof in \cite{PS} that the next such term is
actually of order $R^{-4}$.  However, as this is such a vital point we
give here the most general proof for it. We follow PS and construct
actions that are $(1,1)$ in the na\"ive sense; that is, the net number
of $(+,-)$ indices is $(1,1)$.  We include no terms proportional to
the leading order constraints $\del_\pm X\cdot \del_\pm X$ or to the
leading order equations of motion $\del_{+-} X^\mu$; otherwise they
can be of arbitrary form. Clearly such actions can be constructed out
of skeletal forms of the type
\begin{equation}
\label{skeleton}
\frac{X^{\mu_1}_{s_1,m_1}X^{\mu_2}_{s_2,m_2} \cdots X^{\mu_N}_{s_N,m_N}}{L^M}
\end{equation}
by contracting the Lorentz indices $\mu_1,\mu_2,...,\mu_N$ with the
help of \emph{invariant tensors}, that is, with either
$\eta_{\mu,\nu}$ or $\epsilon_{\mu_1,\mu_2,..,\mu_D}$. Let us consider
the potentially \emph{parity-violating} terms involving the
Levi-Civita symbols later. Here $X^\mu_{s,m}$ stands for $m$
derivatives of type $s=\pm$ acting on $X^\mu$.  The numbers
$\{m_i\},M$ are adjusted to achieve the (na\"ive) $(1,1)$ nature.

The PS lagrangian is not strictly a $(1,1)$ form as can be checked
explicitly. However the PS action, to the desired accuracy, is
invariant under the transformation laws of \er{modtrans}. It is
$(1,1)$ only in the na\"ive sense mentioned above.  The na\"ive
criterion is necessary but not sufficient, thus it suffices to prove
the absence of action terms that are $R^{-3}$ using this criterion.
In fact, it is desirable to have a formulation that is manifestly
covariant.  This will be presented elsewhere \cite{HDPM2}.

Only powers of $L$ have been used in the denominator to get a $(1,1)$
form. It may appear that any scalar in target space would have
sufficed. However, the action should not become \emph{singular} on any
fluctuation. Thus a scalar, say, of the type $\del_+^2 X\cdot\del_-X$
would not be permissible as it vanishes with $Y$. Whatever is in the
denominator must be of the form $\del_+X\cdot\del_-X+\cdots$; this can
always be expanded around the dominant $L$ term to produce forms as in
\er{skeleton}. A covariant formulation \cite{HDPM2} gives a natural
explanation for this as well as for the forms considered in
\er{skeleton}.

All those cases where the Lorentz contractions produce additional
factors of $L$ can be reduced to forms with lower $N$; we therefore
need not consider cases where the number of factors with higher
derivatives ($m\ge 2$) is smaller than the number with only single
derivatives. On the other hand cases with more higher-derivative
factors than single-derivative factors are less dominant. Thus for the
even-$N$ case considered first (taken as $2N$ from now onwards) we
need to consider the maximal case of exactly $N$ single-derivative
terms and $N$ terms with all possible higher derivatives.

Among the single-derivative terms, let $n_+$ be the number with
$+$-derivatives; then there are $N-n_+$ single derivative terms with
$-$-derivatives. Among the higher derivative terms let $p_+$ be the
number of terms with only $+$-derivatives, and likewise $p_-$. Let
$m_+$ be the total number of higher $+$-derivatives and $m_-$ the
corresponding number of higher $-$-derivatives.  As $p_++p_- = N,
m_+\ge 2p_+, m_-\ge 2p_-$, it follows that $m_++m_-\ge 2N$,
$m_++n_+=m_-+N-n_+$, $M = m_++n_+-1$, and subsequently that 
$2m_+\ge 3N - 2n_+$.

Now the leading-order behaviour of such a term is
$R^{N-2(m_++n_+-1)}$. On noting that $N+2-2n_+-2m_+ \le 2-2N$ we see
that for $N\ge 3$ the leading behaviour of the action is at most
$R^{-4}$. The case $N=2$ is precisely the PS action with $R^{-2}$
behaviour.  The dominant case among the subdominant class for $N=2$
(four factors) is where there are three factors with only higher
derivatives and one with a single derivative which we can take to be
of $+$-type without loss of generality. If $l_+$ denotes the total
number of $+$-derivatives among higher derivatives and likewise $l_-$,
we must have $l_--l_+=1$. As before, if $P_+$ denotes the number of
terms with only $+$-derivatives and likewise $p_-$, we have
$p_++p_-=3$ and then $l_+ \ge 2p_+$, $l_-\ge 2p_-$ and $l_++l_-\ge 6$.
These lead to $l_-^{min} =4,l_+^{min}=3$, giving $M=3$ and the
leading-order behaviour is then of order $R^{-5}$. For $N=1$ (two
factors) we can only have higher-derivative terms and it is easy to
see that the dominant term is $\del_+^2X\cdot\del_-^2X/L$, which in
the context of this analysis is equivalent to the PS action.

Finally we turn to parity-violating cases and first to the case where
there is an \emph{odd} number of $X$ fields present.  This can only
happen when $D$ is odd.  The contraction must be between
$\epsilon_{\mu_1..\mu_{2n+1}}$ and an expression of the form
\begin{equation}
\del_+ X^{\mu_1}\del_- X^{\mu_2}
\del_+^2 X^{\mu_3}\del_-^2 X^{\mu_4}\ldots 
\del_+^{n+1} X^{\mu_{2n+1}}
.\end{equation}
The total number of $+$-derivatives is $n(n+1)/2+n+1$, while the total
number of $-$-derivatives is $n(n+1)/2$. The above expression
multiplied by $\del_-^{n+2}X\cdot\del_+X$ balances the $+,-$
derivatives (terms with $+$ and $-$ interchanged are also allowed).
This has to be divided by $(\del_+X\cdot\del_-X)^{n(n+1)/2+n+1}$,
producing a leading behaviour of $R^{3-n^2-3n-2}$ or
$R^{-(n^2+3n-1)}$. This has the potential $R^{-3}$ behaviour in $D=3$
and less dominant behaviour for higher $D$. In $D=3$ this behaviour is
$R^{-3}\epsilon_{\mu_1\mu_2\mu_3}e_+^{\mu_1}e_-^{\mu_2}\del_+^2
Y^{\mu_3} e_+\cdot\del_-^3 Y$ which can be rewritten by partial
integration as
$-R^{-3}\epsilon_{\mu_1\mu_2\mu_3}e_+^{\mu_1}e_-^{\mu_2}\del_+ Y^{\mu_3} e_+\cdot\del_+\del_-^3 Y$ 
and can therefore be dropped as it is proportional to the 
leading-order equation of motion.

In even dimensions a similar analysis shows that the leading behaviour
is $R^{4-D(D+2)/4}$ which need not be considered for $D\ge 6$. For
$D=4$ this is superficially $R^{-2}$ but again both $R^{-2}$ and
$R^{-3}$ terms are proportional to $\partial_{+-}Y$.

\section{Higher corrections to ground-state energy}
\label{sec_high}
From the expression for the ground-state momentum, it is clear that
all higher corrections are determined by
$k^2+\frac{R}{a^2}(e_++e_-)\cdot k$ (\er{restenergy}) which was only
calculated to leading order in \er{leadingvir}. Thus an order-$R^{-n}$
correction to this would result in order-$R^{-n-1}$ and higher
corrections to the spectrum; here we need to investigate both $R^{-1}$
and $R^{-2}$ corrections. As this quantity is just a sum of the $L_0$
and ${\tilde{L}}_0$ conditions, we need to calculate up to
order-$R^{-2}$ corrections to $L_0$ and ${\tilde L}_0$, or
equivalently to $T_{--}$.

As the transformation laws \eqref{modtrans} have a leading part linear
in $R$, additional terms in the action at order $R^{-3}$ would in
principle have induced $R^{-2}$ corrections to $T_{--}$. That would in
turn have changed the $R^{-3}$ terms in the ground-state energy. This
is the reason why the absence of such terms in the action needs to be
established so carefully. Absence of such terms also means that the
expression for $T_{--}$ in \er{fullT__} can be consistently expanded
to keep order-$R^{-2}$ terms. We give here the \emph{on-shell}
expression to the desired order;
\begin{eqnarray}
\label{highT}
&&T_{--}= -\frac{R}{a^2}e_-\cdot\del_-Y-\frac{1}{2a^2}\del_-Y
\cdot\del_-Y-\frac{\beta}{R}e_+\cdot\del_-^3Y \nn\\
&&-\frac{\beta}{R^2}\big[2(e_+\cdot\del_-^2Y)^2
+2e_+\cdot\del_-^3Y(e_+\cdot\del_-Y+e_-\cdot\del_+Y) \nn\\
&&+2e_-\cdot\del_-^2Ye_-\cdot\del_+^2Y+\del_+Y\cdot\del_-^3Y
\big]
.\end{eqnarray}

We see hence that $L_0$ and ${\tilde{L}}_0$ do not receive any
order-$R^{-1}$ correction.  
At this point, the $T_{--}$ of \er{highT} does not seem \emph{holonomic} as
there are $+$-derivative terms occurring in $T_{--}$,
while the Noether procedure necessarily gives a
$T_{--}$ which satisfies $\del_+ T_{--}=0$. 
The resolution of this apparent contradiction lies in the fact that the solution of the
full equation of motion \eqref{modEOM} can no longer be split into a sum of
holonomic and antiholonomic pieces.

Because of the absence of additional terms in the action with the
leading $R^{-3}$ behaviour, the equation of motion \eqref{modEOM} is
sensible inclusive of $R^{-3}$ terms. Now we expand this expression
and retain terms up to order $R^{-3}$;
\begin{eqnarray}
\label{R3EOM}
&&\frac{2}{a^2}\del_{+-}Y^\mu =
- 4\frac{\beta}{R^2}\del_+^2\del_-^2 Y^\mu \nn\\
&&\quad-4\frac{\beta}{R^3}\bigg[
\del_+^2\big\{
              \del_-^2Y^\mu(e_+\cdot\del_-Y+e_-\cdot\del_+Y)
                \big\} \nn\\
        &&\quad\quad+\del_-^2\big\{
               \del_+^2Y^\mu(e_+\cdot\del_-Y+e_-\cdot\del_+Y)
                 \big\}  \nn\\
&&\quad\quad + 4 e_+^\mu\del_-(\del_+^2\cdot\del_-^2Y)+e_-^\mu\del_+(\del_+^2Y\cdot\del_-^2Y)
\bigg]
.\end{eqnarray}
We can solve this equation iteratively by writing $Y^\mu = Y^\mu_0+Y^\mu_1$ 
where $Y^\mu_0$ is a solution of the leading order equation of motion.
Keeping terms only up to order $R^{-3}$ we obtain an 
expression which can be readily integrated to yield
\begin{eqnarray}
\frac{2}{a^2}\del_{-}Y_1^\mu &=& 4\frac{\beta}{R^3}
\big( e_+^\mu\del_+Y_0\cdot\del_-^3Y_0+e_-^\mu\del_+^2Y_0\cdot\del_-^2Y_0 \nn\\
&-&\del_-^2Y_0^\mu e_-\cdot\del_+^2Y_0-\del_+Y_0^\mu e_+\cdot\del_-^3Y_0 \big) 
.\end{eqnarray}
Examining \er{highT} one sees that to order $R^{-2}$ only the first term linear 
in $R$ contributes additional non-holonomic terms which \emph{exactly} 
cancel the remaining non-holonomic pieces.
This immediately leads to the \emph{manifestly holonomic} representation 
of $T_{--}$ to order $R^{-2}$,
\begin{eqnarray}
T_{--} &=& -\frac{R}{a^2}e_-\cdot\del_-Y_0-\frac{1}{2a^2}\del_-Y_0\cdot
\del_-Y_0 -\frac{\beta}{R}e_+\cdot\del_-^3Y_0 \nn\\
&-&2\frac{\beta}{R^2}e_+\cdot\del_-^3Y_0e_+\cdot\del_-Y_0 
-2\frac{\beta}{R^2}(e_+\cdot\del_-^2Y_0)^2
,\end{eqnarray}
whence we obtain the Virasoro generators with higher-order corrections,
\begin{eqnarray}
\label{vircorr}
L_n &=& \frac{R}{a}e_-\cdot\alpha_n+\frac{1}{2}\sum_{m=-\infty}^\infty:\alpha_{n-m}\alpha_m:
+\frac{\beta_c}{2}\delta_{n} \\
&-&\frac{a \beta_c n^2}{R}e_n 
-\frac{\beta_c a^2 n^2}{R^2}\sum_{m=-\infty}^\infty :e_{n-m}e_m: \nn
,\end{eqnarray}
where $e_n\equiv e_+ \cdot \alpha_n$.

Thus we have established that $L_0$ and ${\tilde{L}}_0$ have no
corrections at either $R^{-1}$ or $R^{-2}$ order. As mentioned
earlier, this means all the terms in the ground state energy and the
excited state energies, inclusive of the order-$R^{-3}$ term, are
identical to those in the Nambu-Goto theory.

\section{Conclusions}
Not only is the Polchinski-Strominger action \cite{PS} the unique
effective first-order action for a consistent conformal theory of long
strings, but as we have carefully shown it is unique up to and
including terms of third order in the inverse string length.  The only
remaining freedom in the action is the substitution of the other
equivalent form of the PS term, which we mentioned in
sec.\ref{sec_absence}, and which does not alter our results.

Furthermore, the spectrum is found to coincide with that of the
Nambu-Goto theory, including third order terms.  This
universality explains why comparisons between potentials and excited
state energies in lattice computations \cite{LW,pushan,kuti,caselle,kuti2} and Nambu-Goto
theory have been favourable in the past
even beyond the universal L\"uscher term \cite{pushandass}
(in the case of the ground state energy),
 despite the inconsistency of the Nambu-Goto string outside the critical dimension.
\\

{\bf Note Added:}
The preprint \cite{Drum} came to our attention during the preparation
of this manuscript. Although this work claims to derive \er{vircorr}
it is marred by errors and missing proofs.  Subtleties regarding the
field redefinition in eqns(2.17-2.19) and how this leads to a
non-trivial energy momentum tensor are ignored.  In fact, as mentioned
previously, field redefinitions must be handled with great care.
We have depended on no such field redefinitions in this paper.

A serious deficiency of \cite{Drum} is the assertion that
eqns(2.7-2.10) give the next leading corrections to the PS action.
Not only is this incorrect, as new terms do appear at order $R^{-4}$,
but the faulty analysis thus does not provide convincing evidence that
there are no new terms at order $R^{-3}$.  The issue of at which
orders correction terms exist should be taken seriously. In
particular, the absence of order-$R^{-3}$ terms is absolutely
essential, as can be seen from several stages of sec.~\ref{sec_high}
above; the explicit proof thereof is a main result of the present
work.

\section*{Acknowledgements}
We would like to thank Pavel Boyko and Olexey Kovalenko for their valuable 
discussions and comments.

\end{document}